# The Effects of Cuboid Particle Scattering on Reflected Light Phase Curves: Insights from Laboratory Data and Theory

Colin D. Hamill[1,2], Alexandria V. Johnson[1], Matt Lodge[3], Peter Gao[4], Rowan Nag[1], Natasha Batalha[5], Duncan A. Christie[6], and Hannah R. Wakeford[3]

[1] Department of Earth, Atmospheric and Planetary Sciences, 550 Stadium Mall Drive, Purdue University, West Lafayette, IN 47907, USA; Colin.Hamill@aas.org  
[2] American Astronomical Society, 1667 K Street NW #800, Washington, DC 20006, USA  
[3] School of Physics, University of Bristol, HH Wills Physics Laboratory, Tyndall Avenue, Bristol BS8 1TL, UK  
[4] Carnegie Science Earth and Planets Laboratory, 5241 Broad Branch Road, NW, Washington, DC 20015, USA  
[5] NASA Ames Research Center, MS 245-3, Moffett Field, CA 94035, USA  
[6] Max Planck Institute for Astronomy, Königstuhl 17, D-69117 Heidelberg, Germany  
Received 2024 September 26; revised 2025 April 23; accepted 2025 May 8; published 2025 July 7

## Abstract

Understanding the optical properties of exoplanet cloud particles is a top priority. Many cloud condensates form as nonspherical particles and their optical properties can be very different from those of spheres. In this study, we focus on KCl particles, which likely form as cuboids in warm ($T = 500$–$1000$ K) exoplanet atmospheres. We compare the phase functions (at 532 nm wavelength) of KCl particles computed with Mie theory, the two-term Henyey–Greenstein (TTHG) approximation, laboratory data, and the discrete dipole approximation (DDA). Mie theory assumes scattering from spheres, while TTHG functions are used to approximate cloud scattering in two-stream radiative transfer models like `PICASO`. Laboratory measurements and DDA allow for a robust understanding of scattering from cuboid and deformed cuboid particle shapes. We input these phase functions into `PICASO` using cloud distributions from the cloud model `Virga`, to determine how different phase functions can impact the reflected-light intensities of the benchmark sub-Neptune, GJ 1214b. Simulated reflected light phase curves of GJ 1214b produced using the TTHG, laboratory, and DDA phase curves differ by less than 3 ppm. Our findings suggest that TTHG phase functions may be useful for approximating the scattering intensity of certain cuboid and irregular particle shapes. Future work should expand upon the wavelengths and particles considered to better determine when scattering approximations, like TTHG, may be useful in lieu of more accurate, but time-consuming laboratory measurements and/or nonspherical scattering theory.

*Unified Astronomy Thesaurus concepts:* Exoplanet atmospheres (487); Mini Neptunes (1063); Atmospheric clouds (2180); Interdisciplinary astronomy (804)

## 1. Introduction

Clouds and hazes are ubiquitous in exoplanet observations, from those of super-Earths to hot Jupiters, and their existence poses unique challenges for understanding the scattering properties and radiative balance of planetary atmospheres (M. S. Marley et al 2013; C. Helling 2019; P. Gao et al. 2021). Their presence is frequently inferred from muted spectral feature amplitudes in transmission spectra (e.g., D. Charbonneau et al. 2002; J. L. Bean et al. 2010; H. A. Knutson et al. 2014; L. Kreidberg et al. 2014; I. J. M. Crossfield 2015; D. K. Sing et al. 2016; N. P. Gibson et al. 2017; G. Bruno et al. 2018; P. C. Thao et al. 2020; J. Lustig-Yaeger et al. 2023) and brightness variations in planetary phase curves (e.g., Kepler-7b; D. Kipping & G. Bakos 2011; J. L. Coughlin & M. López-Morales 2012; B.-O. Demory et al. 2013; D. Angerhausen et al. 2015; L. J. Esteves et al. 2015). In addition, the climate impacts of clouds and hazes can vary significantly depending on their location within an atmosphere, as well as particle composition, size, and shape (M. S. Marley et al. 2013; P. Gao et al. 2021).

The influence of particle shape on exoplanet observations is now being explored in greater detail, such as in the cases of aggregate photochemical hazes (D. Adams et al. 2019; P. Lavvas et al. 2019; M. G. Lodge et al. 2023) and nonspherical cloud particles (e.g., R. Tazaki et al. 2016; R. Tazaki & H. Tanaka 2018; K. Ohno et al. 2020; D. Samra et al. 2020; D. Samra et al. 2022; K. L. Chubb et al. 2024). The shape of exoplanet cloud particles, however, still remains largely unconstrained observationally. This is especially true for the large array of solid-phase condensates found in warm (500 K $< T_{eq} <$ 1000 K) and hot ($T_{eq} >$ 1000 K) exoplanet atmospheres (C. Visscher et al. 2010; C. V. Morley et al. 2012; H. R. Wakeford et al. 2017), which could take on a wide array of shapes depending on their chemical composition and the microphysical processes at work in particle formation, such as fluffy aggregates, cubes, or irregular cuboids (K. Ohno et al. 2020; C. D. Hamill et al. 2024b).

Exoplanet cloud models often approximate light scattering by atmospheric particles with Lorenz–Mie (hereafter referred to as Mie) scattering theory, which describes the electromagnetic scattering of perfectly smooth, homogeneous spheres over an arbitrary ratio of size and wavelength (G. Mie 1908; C. F. Bohren & D. R. Huffman 2008). Information about the Mie scattering phase function is typically encoded in the asymmetry parameter, which is computed from the particle sizes and material refractive indices for a given range of wavelengths. The asymmetry parameter is then used to create a Henyey–Greenstein (HG) phase function. HG phase functions are often used in radiative transfer models to calculate the radiative flux of an atmospheric column. While Mie theory and HG phase functions allow for quick calculations of particle scattering, Earth and planetary science literature has shown that nonspherical particles scatter light in a very different way to their spherical particle equivalents (e.g., J. W. Hovenier et al. 2003; M. I. Mishchenko & L. D. Travis 2003; O. Muñoz et al. 2004; D. D. Dabrowska et al. 2015).

The potassium chloride (KCl) particle scattering intensity at a wavelength of 532 nm has recently been measured in the laboratory for KCl particles ranging in average particle radius from 0.6 to 1.2 $\mu$m (C. D. Hamill et al. 2024b). KCl is a salt that naturally forms cubic crystals during deposition (D. Walker et al. 2004) and has been proposed as a cloud species in warm exoplanet atmospheres (e.g., E. M.-R. Kempton et al. 2011; C. V. Morley et al. 2012, 2013; B. Charnay et al. 2015b; P. Gao & B. Benneke 2018; K. Ohno & S. Okuzumi 2018; K. Ohno et al. 2020; P. Gao et al. 2023). While previous papers assumed that KCl particles were spherical and applied Mie theory to compute their scattering properties, C. D. Hamill et al. (2024b) found that Mie theory overestimates the backscattering intensity of cuboid KCl particles by up to an order of magnitude. However, many exoplanet radiative transfer models approximate Mie scattering with HG phase functions for particles of all shapes, and certain features found in Mie scattering are lost during HG approximation. In this paper, we seek to better understand the potential benefits and shortcomings of using HG scattering approximations for nonspherical particles, specifically cubic particles and irregular cuboids. To better understand the effects of particle shape on scattering, we also use theoretical models of nonspherical particles. These models permit efficient investigation of a variety of particle shapes, which is useful for better contextualizing laboratory measurements and HG approximations and allow for investigation into a larger variety of particle sizes and scattering wavelengths than is feasible with laboratory measurements alone.

There are various methods available to determine the light scattering properties of nonspherical particles, such as discrete dipole approximation (DDA) and the T-Matrix method (B. T. Draine & P. J. Flatau 1994; M. I. Mishchenko et al. 2002; M. A. Yurkin & A. G. Hoekstra 2011; M. I. Mishchenko et al 2017). While highly accurate, the long computational time required for these methods often limits their usability for determining light scattering of entire columns of exoplanet clouds with broad particle size distributions. Recently, however, M. G. Lodge et al. (2023) have shown the potential in using DDA in radiative transfer models when the number of dipoles, and thus computation time, is decreased significantly. Another method, the distribution of hollow spheres (DHSs), offers faster computations than DDA by approximating the scattering of basic particle shapes using empty spherical particles (M. Min et al. 2003, 2005). This method has been shown to produce comparable scattering and absorption characteristics to homogeneous nonspherical particles and has been used in theoretical models to determine the scattering characteristics of nonspherical cloud particles in exoplanet atmospheres (D. Samra et al. 2020, 2022; K. L. Chubb et al. 2024). These studies show that nonspherical cloud particles have a slightly increased opacity and albedo compared to spherical particles. Modified mean field (MMF; R. Tazaki & H. Tanaka 2018) theory is an additional useful tool to calculate the scattering properties of nonspherical particles and has been used to investigate aggregate particle structures made up of many individual monomers. The accuracy of MMF theory has been validated against the T-matrix method and it has shown that the optical properties of such particles can differ substantially from spheres (R. Tazaki & H. Tanaka 2018; K. Ohno et al. 2020). While DHS and MMF are useful for understanding the scattering of aggregates or irregularly shaped particles, neither study has been used specifically for cubic or cuboid particles. Thus, they are not well suited to our comparison to laboratory measurements from C. D. Hamill et al. (2024b). Instead, we use DDA following the general methodology of M. G. Lodge et al. (2023), since it offers precise scattering approximations for diversified particle shapes like cubes and irregular cuboids, and can thus be compared to our laboratory data. By comparing laboratory data and DDA models to the HG approximations calculated from Mie theory, we will be able to provide the most robust characterization of its utility in exoplanet radiative transfer models.

In Section 2, we describe the light scattering of Mie theory, HG phase functions, laboratory measurements, and DDA in more detail, and outline the atmospheric and cloud models used in this study. In Section 3, we compare the Mie and HG spherical particle scattering phase functions with the phase functions produced by laboratory measurements and DDA. We then input our phase functions into atmospheric models and contrast the resulting planetary phase curves. Our conclusions are presented in Section 4.

## 2. Methods

### 2.1. Light Scattering Theory

#### 2.1.1. Scattering Intensity from Mie Theory

Mie theory derives an exact solution for the optical properties of homogeneous spheres. Here we outline the key amplitude functions needed to calculate unpolarized scattering intensity with respect to scattering angle. For the full derivation from Maxwell's equations see H. C. van de Hulst (1957). We adopt the same notation as H. C. van de Hulst (1957). Before we can calculate Mie scattering intensity, we must define the Mie coefficients:

$$a_n = \frac{\psi'_n(mx)\psi_n(x) - m\psi_n(mx)\psi'_n(x)}{\psi'_n(mx)\zeta_n(x) - m\psi_n(mx)\zeta'_n(x)}, \quad (1)$$

$$b_n = \frac{m\psi'_n(mx)\psi_n(x) - \psi_n(mx)\psi'_n(x)}{m\psi'_n(mx)\zeta_n(x) - \psi_n(mx)\zeta'_n(x)}, \quad (2)$$

where $m$ is the complex refractive index, $x = 2\pi r/\lambda$ (size parameter), $r$ is particle radius, $\lambda$ is wavelength, and $\psi$ and $\zeta$ are the Ricatti–Bessel functions. Mie's solutions to the amplitude scattering functions are then defined as:

$$S_1(\theta) = \sum_{n=1}^{\infty} \frac{2n+1}{n(n+1)} \{a_n \pi_n(\cos\theta) + b_n \tau_n(\cos\theta)\}, \quad (3)$$

$$S_2(\theta) = \sum_{n=1}^{\infty} \frac{2n+1}{n(n+1)} \{b_n \pi_n(\cos\theta) + a_n \tau_n(\cos\theta)\}, \quad (4)$$

where $\pi_n = P_n^1/\sin\theta$, $\tau_n = dP_n^1/d\theta$, and $P_n^1$ are the Legendre polynomials. The term $\theta$ is the scattering angle from 0° to 180°, such that $\theta = 0°$ represents the forward scattering direction and $\theta = 180°$ represents the backscattering direction. The maximum number of terms needed to calculate the amplitude scattering functions is dependent on the size

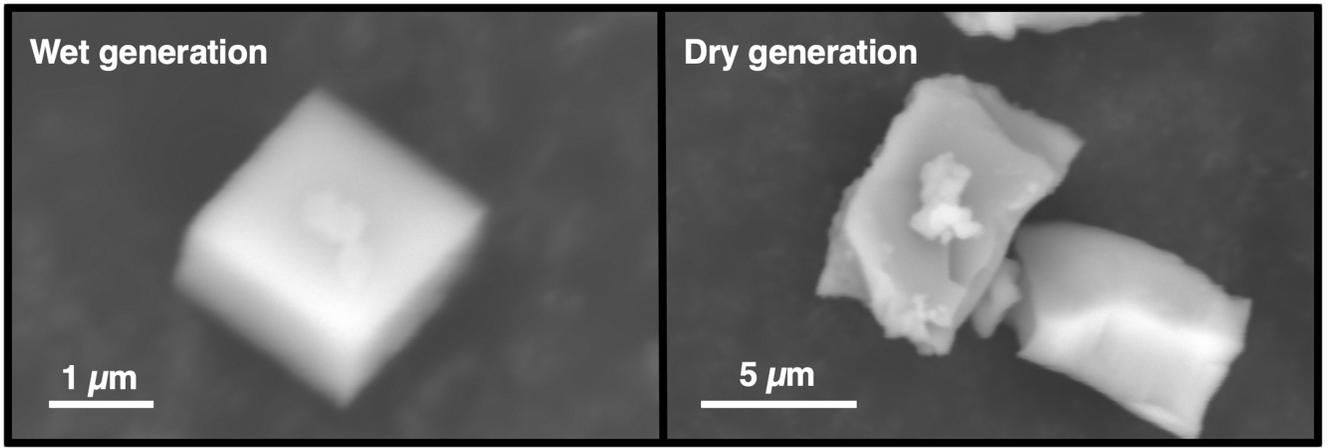

**Figure 1.** Scanning electron microscope (SEM) images of KCl particles created in the laboratory from (left) the wet-generation method used for the small particle size distribution and (right) the dry-generation method used for the medium and large particle size distributions. Images are adapted from C. D. Hamill et al. (2024b).

parameter such that (W. J. Wiscombe 1980):

$$n_{\max} = x + 4x^{1/3} + 2. \quad (5)$$

Once the amplitude scattering functions are known, the first element of the scattering matrix, which describes unpolarized scattering intensity, can be defined by:

$$S_{11} = \frac{1}{2}(|S_1|^2 + |S_2|^2). \quad (6)$$

### 2.1.2. Henyey–Greenstein Phase Functions

Henyey–Greenstein (HG) phase functions approximate the scattering intensity with respect to viewing angle for an ensemble of particles. HG phase functions were first derived from empirical observations to understand the scattering profile of diffuse stellar radiation within the Milky Way with respect to galactic latitude (L. G. Henyey & J. L. Greenstein 1941). All phase functions in this study are normalized using the standard normalization technique such that the area under the phase function is equal to two with respect to the cosine of the scattering angle. One-term HG phase functions are defined as:

$$P_{\text{OTHG}} = \frac{1 - g^2}{(1 + g^2 - 2g\cos\theta)^{3/2}}. \quad (7)$$

The term $g$ is the asymmetry parameter, which is defined as the intensity-weighted mean cosine of the viewing angle (M. I. Mishchenko 1994; C. F. Bohren & D. R. Huffman 2008; K. Ehlers & H. Moosmüller 2023):

$$g = \frac{1}{2}\int_0^\pi P(\theta)\cos\theta \sin\theta d\theta, \quad (8)$$

where $P(\theta) = S_{11}(\theta)/k^2 C_{\text{sca}}$, where $k$ is the wavenumber and $C_{\text{sca}}$ is the scattering cross section (C. F. Bohren & D. R. Huffman 2008). $P(\theta)$ can be calculated for spherical particles using Mie theory, as done in this study, or for irregular particles through laboratory measurements and/or nonspherical scattering theory. The asymmetry parameter $g$ can vary from $-1$ to $+1$. When $g = 1$, light is heavily forward scattered and when $g = -1$, light is heavily backscattered. When $g = 0$, light is equally forward and backscattered and thus represents isotropic scattering.

Although one-term HG phase functions satisfactorily represent the forward scattering peak of particles in the Mie scattering regime, they cannot capture the smaller backscattering peak. To better mimic the scattering profile of real particles, a second term is often introduced:

$$P_{\text{TTHG}} = f\,P_{\text{OTHG}}(\cos\theta, g_f) + (1 - f)P_{\text{OTHG}}(\cos\theta, g_b), \quad (9)$$

where $P_{\text{TTHG}}$ is the two-term HG (TTHG) function, and $g_f$ and $g_b$ are the asymmetry parameters for the forward scattering peak and backscattering peak, respectively. A new term, $f$, describes the fraction of forward scattering compared to backscattering.

### 2.1.3. Laboratory Measurements

Experimental data allows us to test the accuracy of our underlying assumptions and provide further insights about particle scattering than computational methods alone. Since realistic cloud particles are often complicated in shape and size, laboratory measurements can provide a comprehensive view of particle scattering that can be compared with theoretical scattering models and observations from exoplanets. C. D. Hamill et al. (2024b) measured the scattering intensity of three distinct particle size distributions of potassium chloride (KCl) at a wavelength of 532 nm using the Exoplanet Cloud Ensemble Scattering System (ExCESS). ExCESS works by aerosolizing particles of a desired size distribution and composition within a constant flow of nitrogen ($N_2$) gas at room temperature. This system uses two separate aerosolization methods in order to measure different size distributions of particles. The first aerosolization method is a wet-generation method in which a KCl–$H_2O$ solution (0.25 g KCl/50 ml $H_2O$) is aerosolized with an atomizer to produce water droplets that are then desiccated with a drier tube filled with silica gel beads. This wet-generation method produces relatively small ($\bar{r} = 0.6\ \mu$m) particles of a narrow size distribution (weighted standard deviation $\sigma_w = 0.02\ \mu$m) with cubic particle shapes (see Figure 1), as KCl salt naturally forms into cubic crystalline structures during the drying process. The second generation method is a dry-generation method, in which dry KCl particles are ground in a mortar and pestle for a fixed amount of time, in order to produce the desired size distribution. These particles are then agitated and aerosolized with a wrist shaker and flown via $N_2$ carrier gas into the viewing area of ExCESS. This dry-generation method tends to produce a broader particle size distribution with larger particle sizes overall compared to the wet-generation method, and these particles are predominantly shaped like irregular cuboids (see Figure 1) due to the deformation process of mortar-and-pestle grinding. The two dry-generation methods (15 minute grind time and a 0.2 lpm flow rate, and a 5 minute grind time with 0.5 lpm flow rate) used by C. D. Hamill et al. (2024b) produce broad size distributions with a mean particle radius of 0.6 $\mu$m and 1.2 $\mu$m, and weighted standard deviations of 0.2 $\mu$m and 0.5 $\mu$m, respectively. The exact size distributions measured in the laboratory are published by C. D. Hamill et al. (2024b), as measured by an optical particle sizer spectrometer (TSI Optical Particle Sizer 3330) for 16 size bins for particles between ∼0.1 and 5 $\mu$m in radius. Both cubic and irregular cuboid particle shapes are probable in exoplanet atmospheres depending on the formation and deformation processes driven by vertical transport and turbulence.

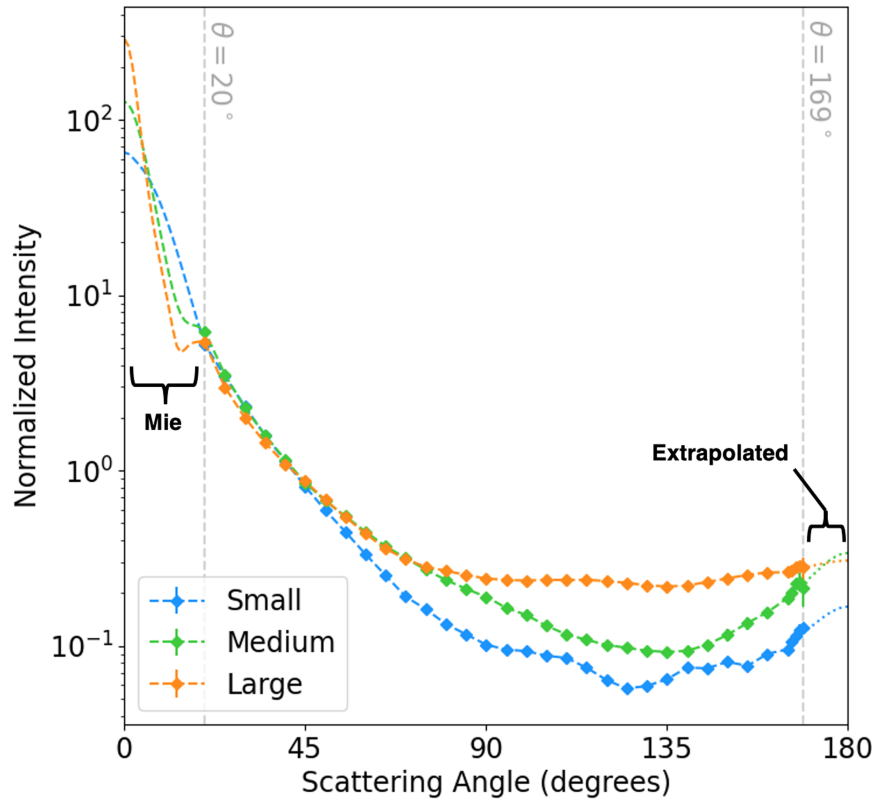

**Figure 2.** The laboratory-measured phase functions (solid) from ExCESS at a wavelength of 532 nm for the small (blue), medium (green), and large (orange) KCl particle size distributions from C. D. Hamill et al. (2024b). The laboratory-measured scattering angles are represented by the diamonds. The error bars represent the standard error of the mean for the experiments and are often smaller than the plotted marker. The minimum and maximum scattering angles ($\theta$) measured in the laboratory are 20° and 169° and denoted by the dashed gray lines. Mie scattering is used to estimate the phase function from 0° to 20° (dashed), while backscattering angles are extrapolated from 169° to 180° (dotted) using a quadratic fit and cubic splines.

As these particles continuously flow into the viewing area and are illuminated by a 532 nm wavelength laser, a photomultiplier tube (PMT) detector sweeps around the plane of illumination and measures the scattering intensity with respect to scattering angle from 20° to 169° in increments of 5°. For very high backscattering angles (165°–169°), measurements are taken in increments of 1°. Measurements at scattering angles less than ∼20° are prohibited by the high intensity of the light potentially damaging the detector. At scattering angles greater than 169°, the PMT detector physically blocks laser light from entering the scattering region. The entire scattering system is enclosed in a dark box to mitigate background noise, and background measurements are taken before each experiment to subtract the background noise at each scattering angle prior to measuring particle scattering. Each phase function is an average of three to five individual experiments. The experimental intensity values are initially normalized such that the intensity at 30° is equal to one, since standard phase function normalization is not possible without full scattering angle coverage.

In order to use the laboratory measurements in radiative transfer modeling, we must extrapolate them from their measured scattering angles (20°–169°) to all scattering angles (0°–180°; Figure 2). To extrapolate the forward scattering peak (0°–20°), we use Mie theory since previous studies showed that the shape and slope of the forward scattering peak is largely independent of particle shape for particles with moderate aspect ratios (i.e., ∼1:2; M. I. Mishchenko et al. 1996, 1997; L. Liu et al. 2003). Mie phase functions are calculated from our measured particle size distributions for KCl with a refractive index of $1.49 + 0i$ (H. H. Li 1976; M. R. Querry 1987), and the results from 0° to 20° are shifted to create a continuous curve with our laboratory measurements. Since these low-angle extrapolations rely on spherical theory to characterize cubic and cuboid particle scattering, however, caution should still be used when interpreting the low-angle intensities shown here. For the backscattering angles, particle shape plays a much larger role in determining scattering intensity and therefore Mie theory cannot be used. Instead, we extrapolate the backscattering angles using a quadratic function (e.g., L. Liu et al. 2003; E. Frattin et al. 2019). We estimate the backscattering intensity at a scattering angle of 180° by using a quadratic equation fit to the laboratory measurements at backscattering angles (135°–169°). Once we have this estimated value of the scattering intensity at 180°, the intensity values from 169° to 180° are produced via a cubic splines interpolation, with the caveat that the slope at 180° must be equal to zero (J. W. Hovenier & D. Guirado 2014). After we extrapolate our laboratory measurements into full phase functions, we renormalize our data from an arbitrary normalization (intensity is equal to 1 at 30°) to the standard normalization technique with respect to cosine of the scattering angle such that the area under the phase function is equal to two. Our extrapolated and normalized laboratory measurements are then placed directly into our radiative transfer model, PICASO (Section 2.2.2), without further modification.

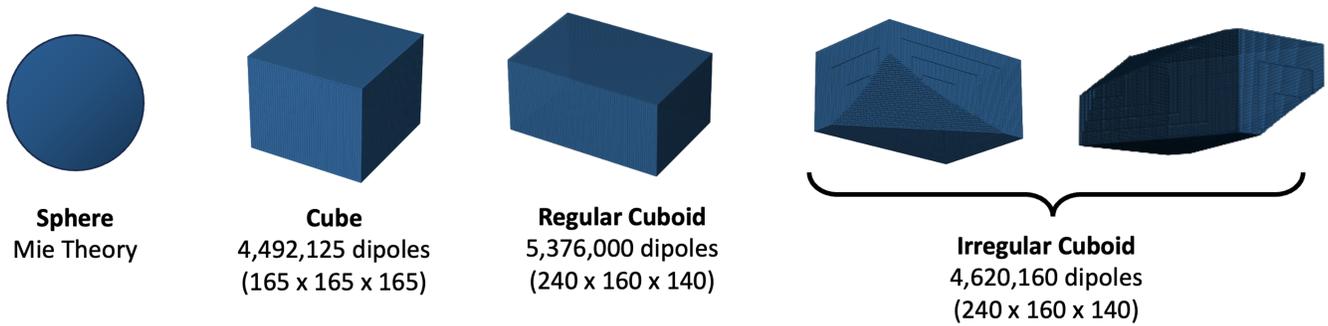

**Figure 3.** The DDA shapes used in this study. We compare scattering intensities from cubes, regular cuboids, and irregular cuboids (using DDA) to those of spheres (using Mie theory). The cubes and cuboids are composed of a series of finite dipoles (numbers of which are shown above, along with the characteristic dimensions of each shape). One single irregular cuboid shape was chosen for use in this study, and views from two different angles are shown here. Dipole widths were chosen so that each shape has identical volume within each respective size distribution bin.

### 2.1.4. Discrete Dipole Approximation (DDA)

The discrete dipole approximation (E. M. Purcell & C. R. Pennypacker 1973; B. T. Draine & P. J. Flatau 1994) calculates scattering from nonspherical particles by dividing them into a series of $N$ small, finite polarizable elements (dipoles), and assessing their combined response to an applied electric field. The theory is exact if $N\rightarrow\infty$, but because the calculation involves solving sets of $3N$ linear equations, in practice the accuracy is limited by how many dipoles can represent the shape with reasonable use of available computing power. For this work, we used the scattering code ADDA v1.4.0 (M. A. Yurkin & A. G. Hoekstra 2011) to determine the phase functions of particles with three distinct geometry types: cubes, rectangular cuboids, and irregular rectangular cuboids (Figure 3).

The dimensions of the shapes were chosen to represent the geometries of realistic KCl particles from C. D. Hamill et al. (2024b) as closely as possible (see the SEM image in Figure 1). The cubes are considered the best representation of the real wet-generation particles (from the small size distribution), and the irregular cuboids were considered the best representation of the dry-generation particles (medium and large size distribution). Regular cuboids were also considered in this study for completeness. We also explored particles with differing levels of surface irregularity/deformation to assess the impact of small and large modifications to the geometry (see the Appendix for details), but we chose a particle shape that best matched the SEM images (by eye) to use in the main study.

All calculations of scattering intensity were averaged over 125 different orientations (changing the direction of incident light), and averages of two perpendicular polarization states. The lattice–dispersion relation (LDR; see D. Gutkowicz-Krusin & B. T. Draine 2004) prescription of DDA was used throughout. All particles were assumed to have a refractive index $m = 1.49 + 0i$, and the size of the dipoles was chosen so that all shapes had equal volume compared to each other, and the spherical particles, using the radii $r$ from each size distribution bin. That is, for a KCl particle composed of $N$ cubic dipoles, each of width $d$:

$$\frac{4}{3}\pi r^3 = Nd^3. \quad (10)$$

The minimum number of dipoles was chosen such that the two key validity conditions for DDA were fulfilled:

(i) the dipole width $d$ must be small compared to the wavelength, such that $2\pi |m| d/\lambda < 0.5$, and
(ii) $d$ must be small enough to describe the geometry satisfactorily.

For this study, condition (i) was achieved by ensuring that all shape files used $N > 4{,}491{,}970$ dipoles. When we changed the particle size, we used the same shape file such that all particles in the distribution were composed of the same number of dipoles. Condition (ii) is more arbitrary, but can be considered fulfilled because the geometry changes are very visibly noticeable at this scale, and the shapes are assumed to have no internal porosity. It is important to state that, when we considered a particular shape type, we assumed that every particle in the distribution has exactly the same geometry (i.e., identical dimensions and deformed in the same way). This is not realistic, but it is a good first step in determining how the results might vary when comparing different shape types, with all other variables held constant.

### 2.2. Atmospheric Models

We apply our nonspherical particle phase function findings to the sub-Neptune GJ 1214b (D. Charbonneau et al. 2009), a warm ($T_{eq} = 500$–600 K) exoplanet with a short period (1.6 days) orbiting an M dwarf (e.g., P. V. Sada et al. 2010; K. B. W. Harpsøe et al. 2013; R. Cloutier et al. 2021). We use three-dimensional (3D) general circulation models (GCMs) of GJ 1214b to map the temperature and chemical composition of the planet with respect to latitude, longitude, and pressure. We then use these quantities to model vertical cloud distributions with the cloud model, *Virga*, and output reflected light phase curves with the radiative transfer code, PICASO. GJ 1214b has been observed to have notably flat transmission spectra (Z. K. Berta et al. 2012; J. D. Fraine et al. 2013; L. Kreidberg et al. 2014), which may be due to an opaque cloud layer in the atmosphere (E. M.-R. Kempton et al. 2011; C. V. Morley et al. 2012, 2013; B. Charnay et al. 2015a; P. Gao & B. Benneke 2018;

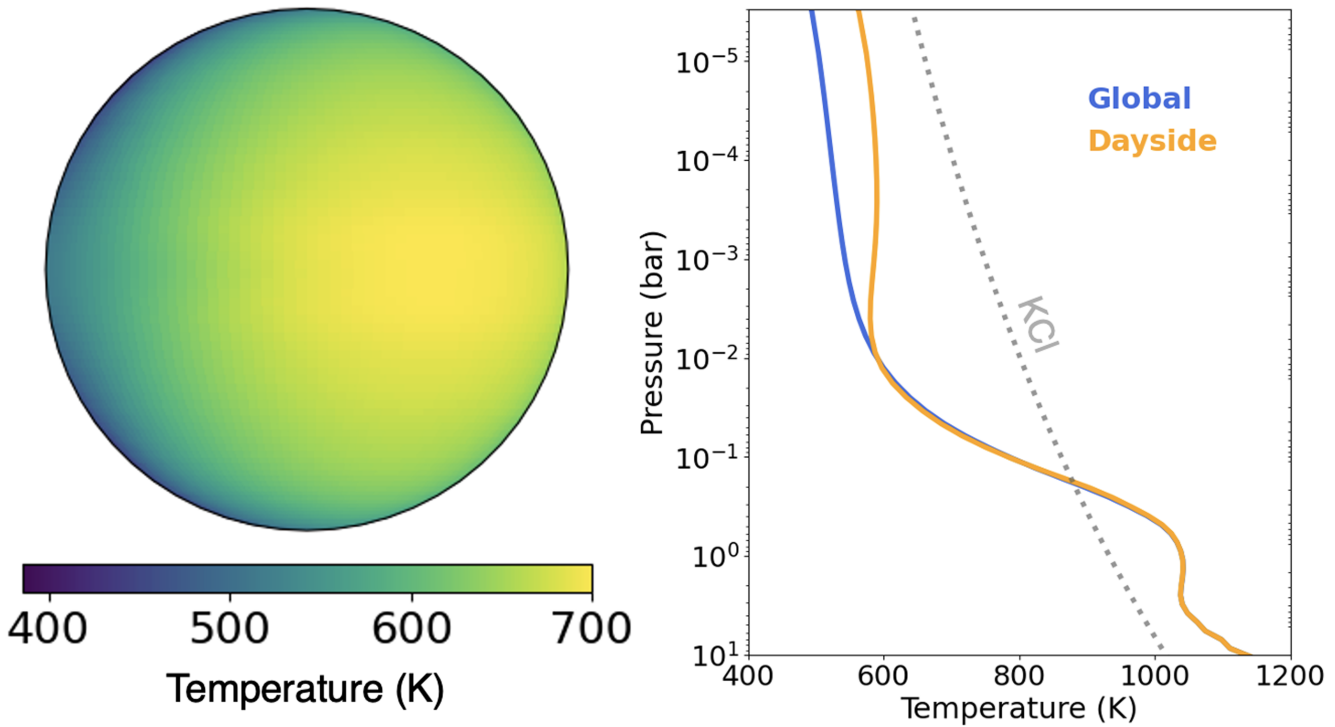

**Figure 4.** (Left) Temperature map of the dayside hemisphere of GJ 1214b (100× solar metallicity) at 1 mbar. (Right) Averaged temperature profiles of GJ 1214b for the dayside (orange) and entire planet (blue), compared to the condensation curve (gray dashed) for KCl for a 100× solar metallicity atmosphere. The GCM is from D. A. Christie et al. (2022) and the KCl condensation curve is from Equation (20) of C. V. Morley et al. (2012).

K. Ohno & S. Okuzumi 2018; K. Ohno et al. 2020; P. Gao et al. 2023). This has been supported by the recent measurement of a full thermal phase curve by the James Webb Space Telescope (JWST) that indicates a high-metallicity (>100× Solar) atmosphere with a thick, reflective layer of clouds or haze (E. M.-R. Kempton et al. 2023). GJ 1214b is expected to form KCl clouds under thermochemical equilibrium with a direct phase change from gas to solid at warm temperatures and pressure levels around $10^{-2}$–$10^0$ bar (K. Lodders 1999; C. V. Morley et al. 2012). While the mean particle radii output by cloud models can vary depending on the atmospheric parameters chosen (i.e., metallicity, sedimentation efficiency), the mean particle radii considered by this study ($\bar{r} = 0.6$–$1.2$ μm) are analogous to the sizes expected in warm exoplanets, especially when $f_{sed} \sim 1$.

### 2.2.1. General Circulation Model

The GCM for this study comes from the Unified Model (UM) from D. A. Christie et al. (2022). The UM GCM solves the deep-atmosphere, nonhydrostatic Navier–Stokes equations and a coupled radiative transfer scheme for mini-Neptunes and hot Jupiters (e.g., D. S. Amundsen et al. 2014, 2017; N. J. Mayne et al. 2014, 2017, 2019; N. Wood et al. 2014; B. Drummond et al. 2018; D. A. Christie et al. 2021, 2022). We use the 100× solar metallicity, clear-sky model of GJ 1214b (Figure 4) from Christie et al. (2022). Although D. A. Christie et al. (2022) also produced GCMs with cloud radiative feedback enabled, they considered the radiative effects of both KCl and ZnS while our study considers only the effect of KCl scattering, and thus we use the clear-sky GCM for our study. Using a clear-sky GCM to study the observational impacts of exoplanet clouds is common (e.g., D. Adams et al. 2022; C. D. Hamill et al. 2024a) and further justified because we are interested in comparing the effects of different scattering approximations in PICASO rather than analyzing the planetary climate of GJ 1214b. We focus on the 100× solar metallicity GCM because multiple studies have shown that GJ 1214b's atmosphere requires high metallicity (≥100× solar) in order to match or come close to the observations (e.g., C. V. Morley et al. 2015; K. Ohno & S. Okuzumi 2018; K. Ohno et al. 2020; D. A. Christie et al. 2022; P. Gao et al. 2023; E. M.-R. Kempton et al. 2023). In addition, we explored the 1× solar metallicity case and found similar results (see the Appendix). While atmospheric metallicities greater than 100× are preferred for GJ 1214b (e.g., P. Gao et al. 2023; E. M.-R. Kempton et al. 2023), our goal here is not to reproduce the planetary phase curve but rather to understand how our scattering phase functions affect it.

### 2.2.2. PICASO

PICASO is an open-source radiative transfer model that computes the reflected light, transmission, and thermal emission of exoplanets in one or three dimensions (N. E. Batalha et al. 2019; N. Batalha & C. Rooney 2020; N. Batalha et al. 2022). For a full description of the code heritage and functionality used in this paper, please see N. E. Batalha et al. (2019), D. Adams et al. (2022), N. Robbins-Blanch et al. (2022), and C. D. Hamill et al. (2024a). To output relative-flux reflected light calculations for an exoplanet, the following parameters are needed: planetary mass, radius, semimajor axis, and pressure/temperature profile(s), as well as the stellar mass, radius, flux, and metallicity (Table 1). To extend the reflected light calculations to a full phase curve, we use the pressure/temperature and $K_{ZZ}$ profiles of the atmosphere with respect to longitude and latitude supplied by the UM GCM described above (Figure 4). The chemical composition of each pressure/temperature profile is calculated assuming thermochemical

**Table 1**
Properties of the GJ 1214b System

| Parameter | Value |
|---|---|
| $M_P$ ($M_J$) | 0.02 |
| $R_P$ ($R_J$) | 0.254 |
| $T_*$ (K) | 3026 |
| $R_*$ ($R_\odot$) | 0.216 |
| $a$ (au) | 0.014 |
| [Fe/H] | 0.39 |
| $\log g$ (cgs) | 4.94 |

**Note.** From K. B. W. Harpsøe et al. (2013).

equilibrium (S. Gordon & B. J. Mcbride 1994; C. Visscher et al. 2010; C. Visscher & J. I. Moses 2011; M. S. Marley et al. 2021). Gas opacities come from the Resampled Opacity Database for PICASO v2 (N. Batalha et al. 2020a).

PICASO is coupled to Virga (Section 2.2.3), allowing easy integration of Virga's cloud optical properties into PICASO's scattering approximations using TTHG functions (N. Batalha 2020). PICASO uses HG/TTHG functions for the direct scattering component of radiative transfer only. This study does not consider multiple scattering since it will have a negligible impact when particle asymmetry parameters are below ∼ 0.8–0.9 (T. D. Robinson 2017). While the TTHG parameters ($g_f$, $g_b$, and $f$) vary depending on particle size, shape, and composition, PICASO prescribes a set relationship between these three parameters to estimate particle scattering from a variety of atmospheric particles. Since this study is comparing various scattering approximations to PICASO specifically, we parameterize $g_f$, $g_b$, and $f$ in the same way as PICASO's default method (N. E. Batalha et al. 2019), as adopted from K. L. Cahoy et al. (2010), such that $g_f = g$, $g_b = -g/2$, and $f = 1 - g_b^2$, where $g$ is the asymmetry parameter calculated directly from the Mie phase functions from C. D. Hamill et al. (2024b) using Equation (8) These relations between $g_f$, $g_b$, and $f$ were created to represent the "very strong forward scattering and moderate backscattering" of atmospheric clouds, and were informed by observations of the solar system gas giants (K. L. Cahoy et al. 2010). All phase functions are normalized such that the area under the phase function is equal to two with respect to the cosine of the scattering angle. Normalizing in this manner also matches the method found in PICASO for its native HG phase functions, and ensures that intensities are all normalized to the expected values inherent to PICASO.

Reflected light phase curves are computed in PICASO by calculating the intensity at multiple plane-parallel points across the dayside hemisphere. PICASO then uses the Chebyshev–Gauss integration method to integrate over all intensities. We consider 15 discrete phase angles in PICASO to represent our phase curve, following the methods of N. Robbins-Blanch et al. (2022) and C. D. Hamill et al. (2024a; Figure 5). We use a 20 × 20 spatial grid to represent the visible dayside hemisphere at each discrete phase angle, as C. D. Hamill et al. (2024a) found this spatial resolution offers an optimal balance between resolution and computation time.

### 2.2.3. Virga

Virga is an open-source cloud model, based on *EddySed*, that computes the optical depths, single-scattering albedo, and asymmetry parameters of spherical condensates under phase-equilibrium as a function of atmospheric pressure and wavelength (A. S. Ackerman & M. S. Marley 2001; N. Batalha et al. 2020b). The KCl condensate refractive indices are from M. R. Querry (1987) and input into Virga by N. Batalha & M. Marley (2020). Virga controls the vertical extent and particle size distribution of clouds by balancing upward convection, parameterized by the vertical eddy diffusion coefficient ($K_{zz}$), with downward sedimentation, such that:

$$-K_{zz}\frac{\partial q_t}{\partial z} = f_{\text{sed}} w_* q_c, \quad (11)$$

where $z$ is the atmospheric height, $q_t$ is the total mixing ratio of condensate and vapor, $q_c$ is the mixing ratio of condensates, $f_{\text{sed}}$ is the sedimentation efficiency, and $w_*$ is the convective velocity scale. We adopt $K_{zz}$ values from B. Charnay et al. (2015a), who derived one-dimensional $K_{zz}$ profiles to describe the global average mixing of GJ 1214b given by the function:

$$K_{ZZ} = \frac{K_{ZZ,0}}{P_{\text{bar}}^{0.4}}, \quad (12)$$

where $P$ is atmospheric pressure (bars) and $K_{ZZ,0}$ is equal to $7 \times 10^2$ and $3 \times 10^3$ m$^2$ s$^{-1}$ for the 1× and 100× solar metallicity atmospheres, respectively. We use this $K_{zz}$ profile at every latitude/longitude grid point in our GCM, as was done in D. A. Christie et al. (2021, 2022). The sedimentation efficiency, $f_{\text{sed}}$, is a tunable parameter set by the user that controls the balance between upwelling and sedimentation of condensate particles. Following previous studies, which suggest that GJ 1214b's sedimentation efficiency must be low ($f_{\text{sed}} \leqslant 0.1$; C. V. Morley et al. 2013, 2015; D. A. Christie et al. 2022) to match observations, we choose to set $f_{\text{sed}} = 0.1$ for this study. We also explored cases where sedimentation efficiency is set to 1.0 (see the Appendix).

We use Virga to compute single-scattering albedo and extinction as a function of pressure and wavelength. The asymmetry parameters output by Virga are no longer taken into consideration. Instead we input our DDA, lab, or TTHG phase functions directly into PICASO's reflected intensity functions. This means our Virga cloud vertical distributions are not self-consistent with the particle size, since we are inputting phase functions created from different particle size distributions than those computed by Virga. Due to limitations in laboratory aerosolization, our phase functions only consider KCl particles from ∼0.1 to 5 $\mu$m particle radius, which is a narrower size distribution than those considered in Virga with comparable particle radii. As such, while our findings offer an understanding of how particle shape and phase function approximation can impact the reflected light from cloudy exoplanets when all other variables are held constant, caution should be used when comparing directly to future reflected light observations of GJ 1214b or other particle compositions.

## 3. Results

### 3.1. Comparing Mie, TTHG, DDA, and Laboratory Phase Functions

We present the Mie, TTHG, extrapolated laboratory, and DDA scattering intensities with respect to scattering angle for a wavelength of 532 nm in Figure 6. Our extrapolated laboratory measurements and DDA phase functions are placed directly into `PICASO`. All phase functions are normalized using the same method such that the area under the curve is equal to two with respect to the cosine of the scattering angle. Our phase functions are calculated for three distinct particle size distributions labeled as small, medium, and large. The small size distribution has a mean particle radius of 0.6 μm with a narrow weighted standard deviation of 0.02 μm, the medium size distribution has the same mean particle size with an order of magnitude larger weighted standard deviation of 0.2 μm, and the large size distribution has a mean particle radius of 1.2 μm with a weighted standard deviation of 0.5 μm.

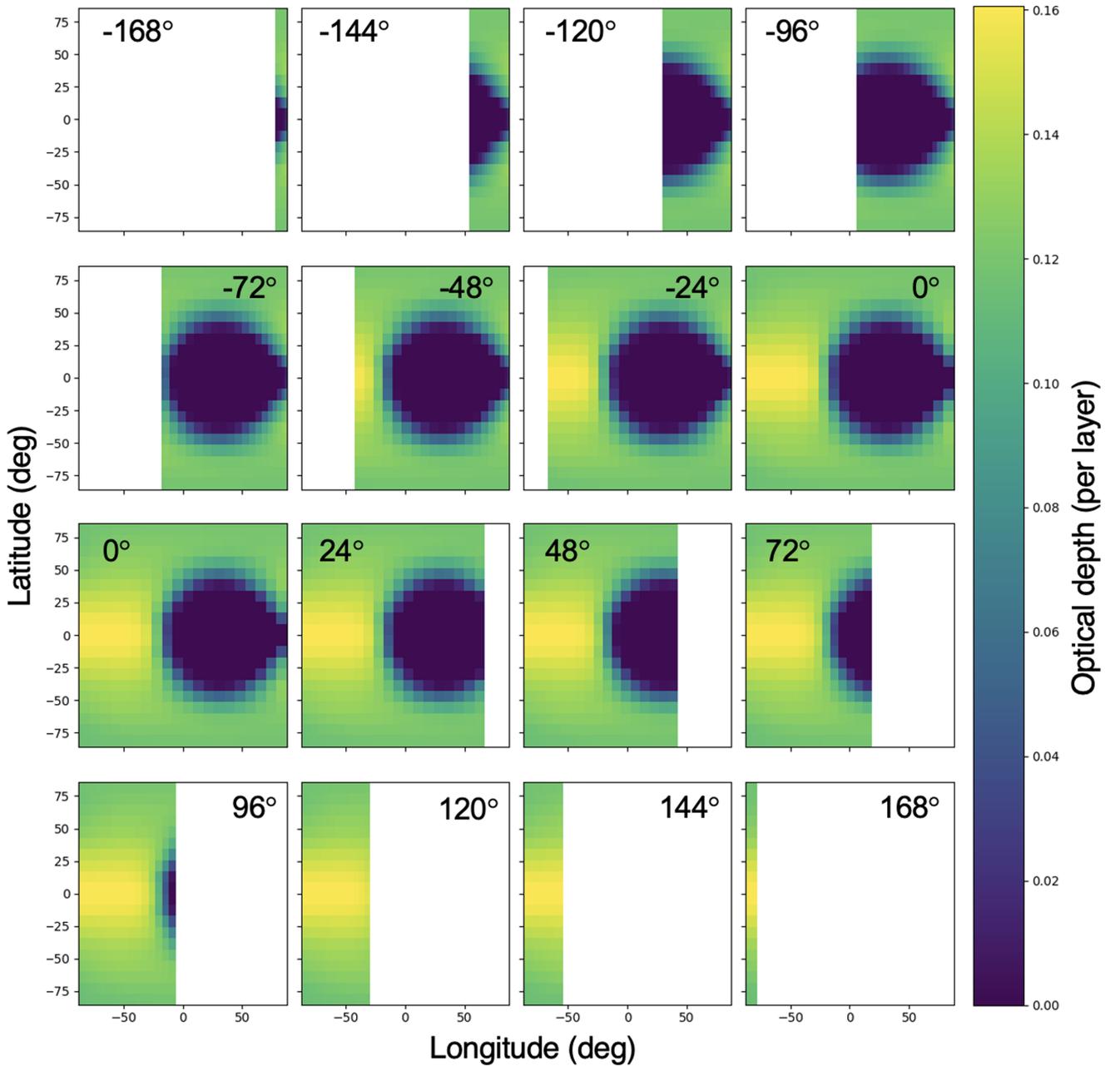

**Figure 5.** KCl cloud optical depth maps of the visible dayside hemisphere at 1 mbar and a wavelength of 532 nm for our GJ 1214b GCM (100× Solar metallicity) at 16 phase angles (−168° to 168°). KCl cloud optical depth is computed with `Virga` with $f_{sed} = 0.1$. The colored part of each map represents the part of GJ 1214b's dayside visible to an observer on Earth, resolved by a 20 × 20 spatial grid. Phase angles of 0° are shown twice and represent the secondary eclipse. A phase angle of ±180° (not shown) represents the primary eclipse (transit) where no reflected light is visible to the observer. The substellar point is at the center of each map (0° latitude, 0° longitude).

The Mie phase functions (gray dotted) are originally calculated by C. D. Hamill et al. (2024b) and show the exact scattering solution assuming the KCl particles are homogeneous spheres with a refractive index of $1.49 + 0i$. The Mie phase functions show many undulations and a steep backscattering (>135°) slope for all three size distributions. The small, medium, and large Mie phase functions have asymmetry parameter values of 0.637, 0.714, and 0.737, respectively. The TTHG phase functions (black) are calculated using Equation (9) and exhibit a heavily reduced backscattering peak compared to Mie backscattering, a characteristic that is common for TTHG phase functions (S. Sanghavi et al. 2021). The TTHG phase functions also exhibit less forward scattering, and tend to be brighter in the mid (45° to 135°) viewing angles compared to the Mie phase functions.

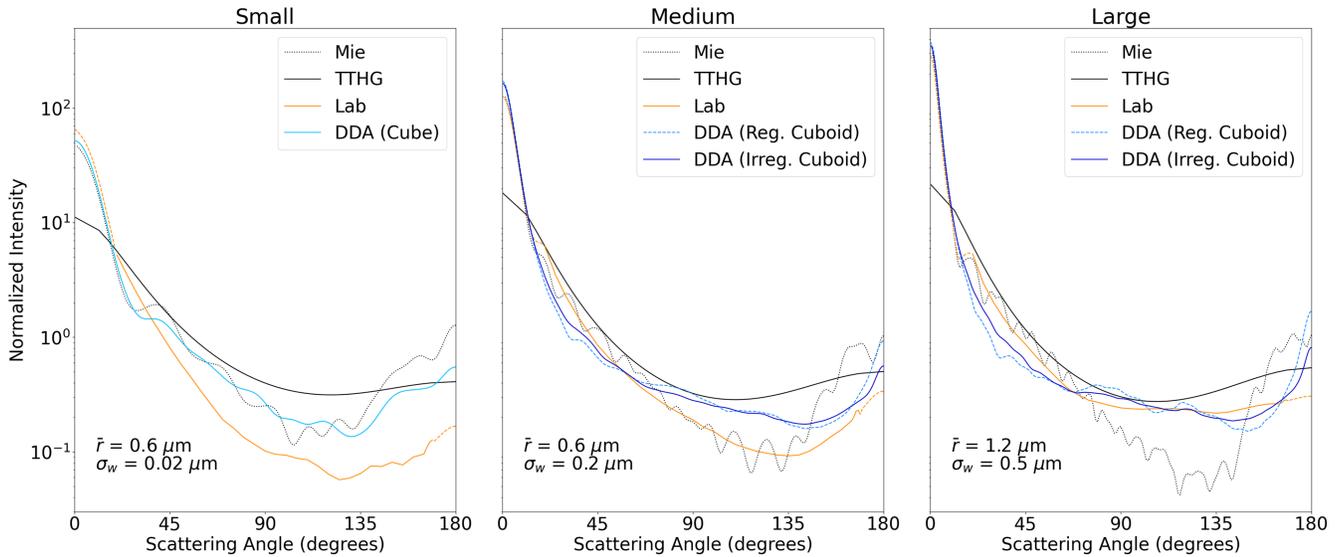

**Figure 6.** All phase functions considered in this study for the small (left), medium (middle), and large (right) KCl particle size distributions at a wavelength of 532 nm. The mean particle radii and weighted standard deviation of each size distribution is given at the bottom left of each plot. The TTHG phase functions (black) are computed using the same equation used in PICASO, with the Mie phase functions (gray dotted) as input. Laboratory phase functions (orange) are extrapolated to all scattering angles and stem from the scattering of light by ensembles of cubic or irregular cuboid particles. The orange dashed lines show the regions of extrapolation of our laboratory data at low and high scattering angles. The small size distribution is produced via KCl desiccation and yields cubic particles, while the medium and large size distributions are produced via dry particle aerosolization and yield irregular cuboids. Phase functions computed from DDA are shown for cubic (sky blue), regular cuboid (blue dashed), and irregular cuboids (dark blue). The Mie and lab phase functions shown here are from C. D. Hamill et al. (2024b).

The extrapolated laboratory phase functions are shown in orange (Figure 6). Due to the cubic and irregular cuboid particle shapes created in the laboratory, these phase functions show a flatter backscattering slope compared to the Mie phase functions. The laboratory phase functions are dimmer in the mid viewing angles for the small particle size distribution compared to their Mie or TTHG phase functions. The dim backscattering peak of the laboratory phase functions are best matched by the TTHG phase functions. In other words, the dimness of the backscattering signal inherent to TTHG phase functions, which makes TTHG a less-than-ideal choice for approximating spherical particle scattering, provides a closer approximation to cubic/cuboid particle scattering.

The DDA phase functions for cubes (sky blue), regular cuboids (blue dashed), and irregular cuboids (dark blue) are also shown in Figure 6. For the small size distribution, DDA is somewhat more successful at reproducing the shape of the laboratory data than both Mie theory and the TTHG—it correctly characterizes the local minima features at ∼125° and predicts a similar backscattering slope at high angles. The DDA phase function also reproduces the smoothness of the lab data, whereas the curve predicted by Mie theory has many resonance features that do not match the experimental observations. However, the intensities predicted by DDA are higher than those determined in the lab; this could be due to slight alignment errors in the experimental setup or extinction effects caused by ambient particles in between the PMT detector and the scattering region, causing slightly dimmer measured intensities than expected. For the medium and large size distributions, the cuboid and irregular cuboid DDA phase functions show less variation in intensity compared to their respective Mie phase functions at mid to high scattering angles. As expected, the backscattering peak of the DDA phase functions is lower than the Mie phase functions, except for the regular cuboid in the large size distribution, and dependent on particle shape. The cubic particle shapes of the small size distribution produce a backscattering peak that is over two times lower than the corresponding Mie phase function. As seen in the medium and large size distributions, irregular cuboids also lead to lower backscattering peaks by up to a factor of 2 compared to regular cuboids, as well as slightly less variation in the phase function at mid scattering angles than the regular cuboids (Figure 6). We explored other particle shapes with DDA with varied cuboid irregularity and found similar results in the resultant phase functions (see the Appendix).

The DDA phase functions are smoother than the Mie phase functions because the resonance features that occur in perfect spheres (because of their symmetrical geometry) do not occur in the same way for irregular shapes. While DDA for rectangular cuboids predicts two minor resonance features that are not present in the lab data, these are noticeably smaller for the irregular cuboids. In addition we note that the existence of the resonance features is likely caused by the assumption that all particles were the same shape in this model. If a truly random ensemble of unique and random irregularities were used to represent each particle, these resonances would likely disappear and produce an even smoother phase function, better matching the lab data. However, this would require calculating the optical properties of each shape uniquely, which becomes increasingly difficult depending on how many particles are modeled due to computational limitations.

### 3.2. Single-wavelength Phase Curves with PICASO

The discrete wavelength (532 nm) reflected light phase curves of GJ 1214b computed using our measured and theoretical phase curves, assuming a 100× solar metallicity atmosphere and a KCl particle sedimentation efficiency of 0.1, are shown in Figure 7. Every phase curve uses the same cloud locations, single-scattering albedo, and extinction coefficients as output from Virga and thus differences between each phase curve are the result of the different particle scattering phase functions. For each of the three particle size distributions explored in this study (small, medium, and large), Figure 7 shows 532 nm phase curves using either TTHG (black), laboratory (orange), or DDA (blue) phase functions, where cubic DDA particles are sky blue for the small size distribution, and regular and irregular cuboids are shown as blue dashed or dark blue solid lines, respectively, for the medium and large particle size distributions. The plots in the bottom row of Figure 7 show the relative-flux difference between the Lab and TTHG data (orange) or between the DDA and TTHG data (blue).

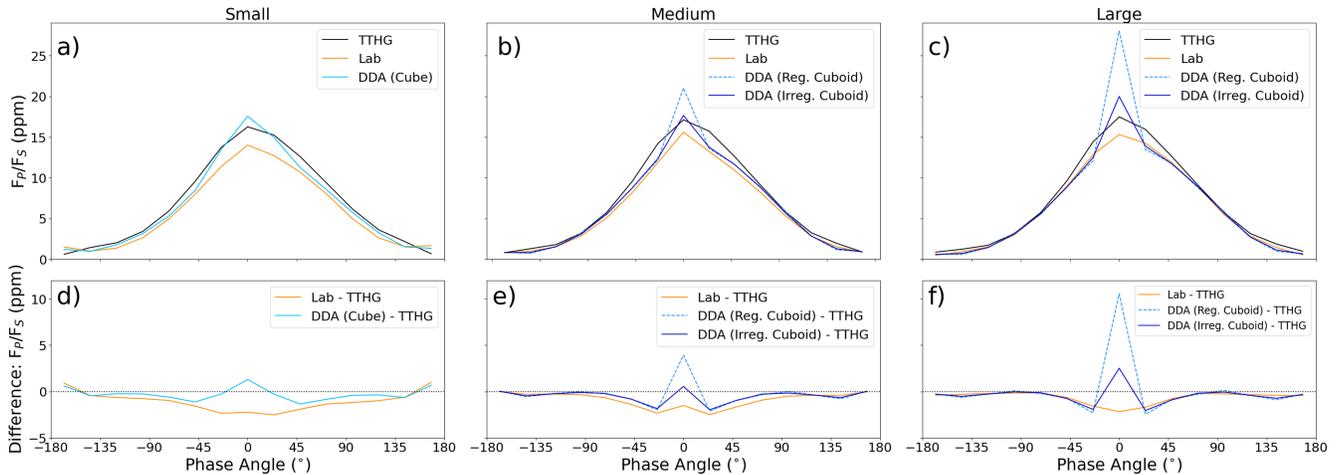

**Figure 7.** Reflected light phase curves of GJ 1214b (100× solar metallicity) at 532 nm showing the planet-to-star flux ratio in parts per million (ppm) as a function of phase angle (−168°–168°) for the (a) small, (b) medium, and (c) large particle size distributions with the TTHG (black), laboratory (orange), cubic DDA (sky blue), regular cuboid DDA (blue dashed), or irregular cuboid DDA (dark blue) phase functions used as input. The differences in relative flux between the phase curves are also shown for the (d) small, (e) medium, and (f) large particle size distributions.

The TTHG phase curves represent the base method that PICASO uses to approximate particle scattering by spherical cloud particles. The TTHG phase curves are smooth, due to the smoothness of the TTHG phase functions, and peak at a relative flux of 16–18 ppm for all three size distributions. There is a minor phase offset for all phase curves in Figure 7 caused by the uneven cloud distribution of GJ 1214b's dayside, such that the KCl cloud optical thickness is higher on the western dayside than the hotter eastern dayside (Figure 5).

The phase curves for the laboratory-measured phase functions (orange) are also relatively smooth. These phase curves are dimmer than the TTHG phase curves by 0.1–2 ppm at all phase angles due to the lower intensities of the laboratory phase functions at all scattering angles greater than ∼25°. While the laboratory phase functions are brighter than the TTHG at very low (<25°) scattering angles, these differences only play a role when the planet is very close to the primary eclipse (phase angle of ±180°) and thus the reflected intensity is low or nonexistent.

The reflected intensities derived from the cubic DDA phase function (sky blue) shown in Figure 7(a) are also dimmer than the TTHG phase curve by up to 1 ppm for the majority of phase angles, except for near the primary and secondary eclipse when the DDA phase curve is slightly brighter than the TTHG curve by 1 ppm. The peak in reflected light at secondary eclipse is due to the increased backscattering intensity of the cubic DDA phase function at a scattering angle of 180° (Figure 6). The DDA phase curves for the medium and large size distributions exhibit similar peaks in intensity near secondary eclipse. The regular cuboid DDA phase curves show a maximum difference of 4 and 10 ppm for the medium and large size distributions, respectively, compared to the TTHG curve, while the irregular cuboid DDA phase curves only show a maximum difference of 2–3 ppm for the medium and large size distributions. Observationally distinguishing between cubes or irregular cuboids would be unlikely, as current observational uncertainties are higher than the differences shown here. For example, phase curve amplitude errors are >±3 ppm for all 14 hot Jupiters observed with Kepler (420–900 nm) by L. J. Esteves et al. (2015), except for the very dimmest (<5 ppm amplitude) planet observed, TrES-2b. B.-O. Demory et al. (2013) analyzed Kepler data of Kepler-7b and found a phase curve amplitude of 50 ± 2 ppm and an occultation depth of 48 ± 3 ppm. Kepler-7b is an ideal exoplanet for reflected light characterization since it has a high albedo and closely orbits a G-type star. With JWST, which observes at longer wavelengths than considered here, M. Holmberg & N. Madhusudhan (2023) found that contamination for hot Jupiter observations of WASP-39b and WASP-96b can produce errors >100 ppm over the broad wavelength range covered by the Near-Infrared Imager and Slitless Spectrograph (0.6–2.8 $\mu$m). V. S. Meadows et al. (2023) found errors >10 ppm in theoretical transmission spectra of TRAPPIST-1e with a cloudy, Earth-like atmosphere for wavelengths of ∼0.6–1 $\mu$m, even with 100 observed transits.

## 4. Discussion

Our results from Section 3.2 show that the differences in reflected planetary albedo when using more realistic scattering phase functions are small (<3 ppm) compared to the albedos derived from the nominal TTHG functions used in PICASO. This finding suggests that phase-resolved planetary albedo is rather insensitive to the functional form of the scattering phase function, especially at phase angles near primary eclipse (∼±180°) or in quadrature (∼90/270°). Even large, near order-of-magnitude differences in the scattering phase function (Figure 6) lead to only ∼1 ppm differences in reflected albedo for GJ 1214b (Figure 7). The largest differences in albedo occur near secondary eclipse (phase angle = 0°), when the full dayside is visible to the observer, though these yield observational differences of only ∼2–3 ppm. These results imply that the current form of the TTHG phase function, first used by K. L. Cahoy et al. (2010), is adequately approximating the overall scattering profile of cubic and irregular cloud particles given the observational constraints of our current space telescopes (e.g., B.-O. Demory et al. 2013; L. J. Esteves et al. 2015; M. Holmberg & N. Madhusudhan 2023; V. S. Meadows et al. 2023).

Our DDA phase functions (Figure 6) and subsequent reflected phase curves (Figure 7) show that planetary albedo can be up to ~10 ppm greater at secondary eclipse if we assume regular cuboid scattering rather than a TTHG approximation. This implies that changes in planetary albedo caused by certain particle shapes may be more prominent at secondary eclipse. However, it is unlikely that perfect rectangular cuboid particles will form in abundance for KCl or any other cloud species. Furthermore, reflected observations at secondary eclipse are often obstructed by the planet's host star, so it will be difficult to measure the planetary albedo directly at this critical phase angle.

There are limitations and caveats to highlight when considering the takeaways of this study. First, this study is limited to one discrete wavelength (532 nm), one particle composition (KCl), and only three particle size distributions, all with a mean particle radius $\sim 1\ \mu$m. A study exploring a wider parameter space would allow us to better understand the utility and shortcomings of the TTHG phase functions used in `PICASO` across a range of observable wavelengths and cloud scenarios. Second, as we explain in Section 2.2.3, we do not change the single-scattering albedo or extinction parameters calculated by `Virga` for assumedly spherical particles. We hold these values constant in order to explicitly understand how scattering phase functions alter the reflected albedo of GJ 1214b. If we were to change the single-scattering albedo or extinction parameters for the particle shapes used, the output reflected albedo would likely change slightly from what is published here. Since it would be impractical to calculate nonspherical single-scattering albedo and extinction parameters in a self-consistent way outside of the narrow set of size distributions and particle shapes used in this study, we choose to only consider the effects of the TTHG phase function, as the TTHG parameters ($g_f$, $g_b$, and $f$) can easily be modified if needed to improve the phase-dependent cloud scattering intensities used in `PICASO`.

## 5. Conclusions

We compared the measured scattering phase functions of KCl for three distinct particle size distributions from C. D. Hamill et al. (2024b) to TTHG approximations and DDA calculations of the same particle size distributions and provided an analysis of their similarities and differences. We ran DDA calculations for cubic particles, which are representative of the KCl particles produced in the laboratory via the wet-generation method (the small size distribution), and for regular and irregular cuboids similar to those produced in the lab via the dry-generation method for the medium and large size distributions. We then test the differences that these various scattering approximations and measurements have on the reflected intensities of GJ 1214b, a cloudy exoplanet with a potential layer of cubic or irregular cuboid shaped KCl particles.

The overall differences in reflected light intensity as a result of laboratory measurements and DDA approximations are low (0–3 ppm) compared to the TTHG phase curves, where the phase functions are produced using the same equation native to `PICASO`. The standout cases are the regular cuboid DDA phase curves, which produce differences in reflected light intensity of 4–10 ppm for the medium and large size distributions, though these particle shapes are likely not representative of the true particle shapes measured in the laboratory or present in GJ 1214b's atmosphere. The cubic and irregular cuboid laboratory measurements and DDA calculations, however, produce phase curves that are strikingly similar in shape to the phase curves produced with TTHG approximations.

While it is known that TTHG approximations offer a poor alternative to the more accurate Mie calculations (S. Sanghavi et al. 2021), many radiative transfer models still use them due to their fast computation times. This study, however, suggests that TTHG approximations may inadvertently provide a closer estimation for cloud scattering when particles are expected to be cubic or irregularly shaped rather than spherical, as the low backscattering signal and flatness of the TTHG phase functions match well to the measured and DDA phase functions, and thus produces phase curves of similar reflected intensity (differences of $\pm 3$ ppm). Using TTHG phase functions for the purpose of modeling nonspherical particle scattering may be able to provide fast and reasonable approximations for a larger array of particle sizes and wavelengths.

Laboratory measurements and nonspherical scattering models, such as DDA, are always preferred due to their accuracy, and form the basis for comparison to existing scattering approximations. However, both methods take considerable time and resources and are often limited in the particle sizes, shapes, or wavelengths they investigate for any given experiment. Future studies focused on expanding the wavelengths and particle shapes, sizes, and compositions could shed much needed light on how various scattering approximations may be under- or overestimating reflected intensities from cloudy exoplanets. As the next space telescopes, notably the Nancy Grace Roman Space Telescope and the Habitable Worlds Observatory, begin their campaigns in the 2030s and 2050s, respectively, to observe the reflected light of exoplanets in the visible and near-infrared (N. J. Kasdin et al. 2020; S. R. Vaughan et al. 2023), it is more important than ever to ensure we are accurately modeling the optical properties for particles of all shapes and sizes.

## Acknowledgments

We thank the anonymous reviewers for their helpful comments that improved the manuscript. We acknowledge support from the NASA Exoplanet Research Program (80NSSC23K0041), the Frederick Frank fund, and the Keith Burgess scholarship. We would like to thank Zoë M. Leinhardt for reading the manuscript and providing comments that greatly improved the quality of this work.

*Software*: `picaso` (N. Batalha et al. 2022), `virga` (N. Batalha et al. 2020b), Matplotlib (J. D. Hunter 2007), pickle (G. Van Rossum 2020), bokeh (Bokeh Development Team 2014), NumPy (S. van der Walt et al. 2011), ADDA v1.4.0 (M. A.Yurkin & A. G. Hoestra 2011), Irregulator (https://github.com/mglodge/Irregulator).

## Appendix

### A.1. Particle Irregularity with Discrete Dipole Approximation

For our irregular cuboids, which correspond to the dry-generated particles produced for the medium and large size distribution in the laboratory, we explored several varieties of particle irregularity to better understand how particle shape impacts the scattering phase functions. The four cuboid shapes we created are shown in Figure 8 and were generated using the publicly available code "Irregulator"[7] (M. G. Lodge 2024), which applies two deformation processes to a regular rectangular cuboid. First, it defines two triangular planes at opposite vertices of the cuboid and removes any dipoles external to these planes, "slicing" the corners off. Second, it causes a unique deformation at a random point on each surface of the cuboid and linearly deforms the rest of each surface from the edges to this point. The depth of deformation on each surface and the positions of the triangular "slices" are chosen at random, with the possible range of values (and thus the amount of deformation possible) chosen by the user. Here we examined three examples in addition to the regular cuboid: slight irregularity, medium irregularity, and extreme irregularity. All shape files used in this study are made publicly available.[8]

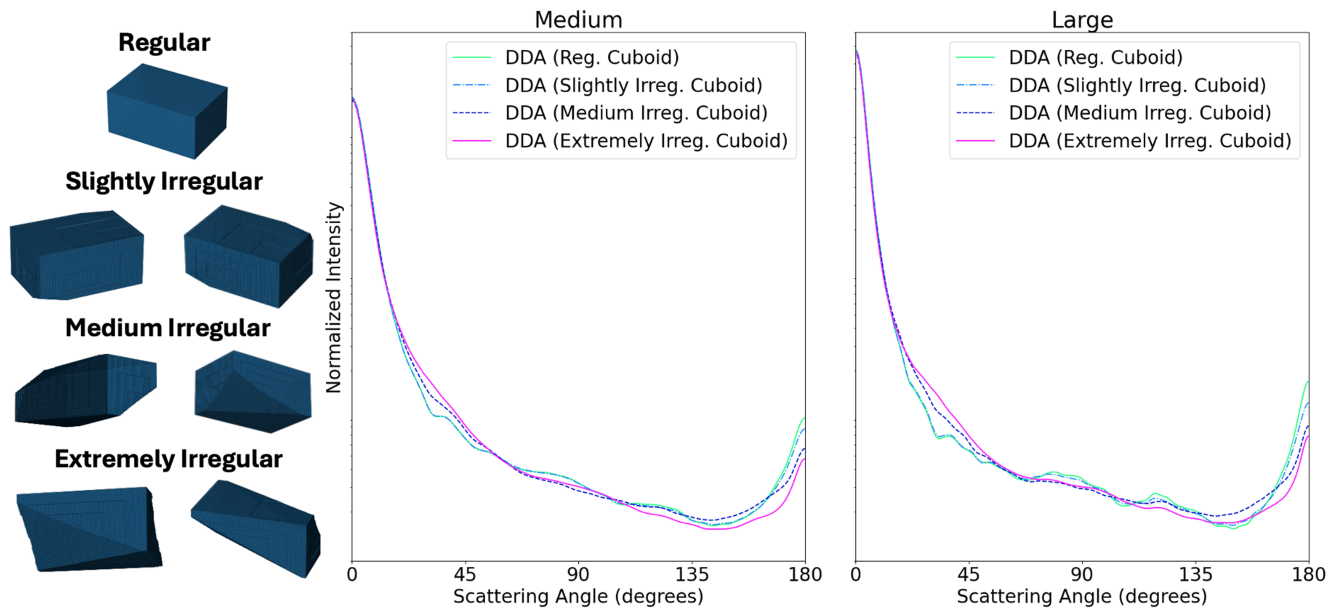

**Figure 8.** (Left) Regular and irregular cuboid shapes used in DDA, with the corresponding phase functions at 532 nm for the medium (middle) and large (right) particle size distributions explored in this study.

The corresponding phase functions produced by each particle shape are shown for both the medium and large size distributions. As the cuboids become increasingly irregular, the phase functions tend to become flatter at mid viewing angles, with less prominent peaks at ∼60° and 110° scattering angles. The backscattering peaks also become smaller as shape irregularity increases. Due to the similarity between the medium irregular cuboids and the shapes of KCl particles seen in SEM images in the laboratory, we choose to focus on this irregular particle shape for the medium and large size distributions, along with the regular cuboid as a point of further comparison.

### A.2. Atmospheric Metallicity and Sedimentation Efficiency

We explored two atmospheric metallicities (1× and 100× solar) and two sedimentation efficiencies (0.1 and 1.0) to better understand how our phase functions altered the observable features of GJ 1214b under a variety of scenarios, using the 1× and 100× solar metallicity, cloud-free GCMs from D. A. Christie et al. (2022) with the associated $K_{ZZ}$ profiles from B. Charnay et al. (2015a). As expected, the overall differences between the laboratory and TTHG phase functions and between the DDA and TTHG phase functions are largely independent of the assumed metallicity or sedimentation efficiency (Figure 9). As an example of this, Figure 9 shows the reflected light phase curves for the TTHG (black), laboratory (orange), and cubic DDA (sky blue) phase functions for four different atmospheres: 1× solar metallicity and $f_{sed}$ = 0.1 (highest line opacity), 1× solar metallicity and $f_{sed}$ = 1.0, 100× solar metallicity and $f_{sed}$ = 0.1 (the case explored in this paper), and 100× solar metallicity and $f_{sed}$ = 1.0 (lowest line opacity). The brightness differences in the four atmospheric scenarios are due to differences in cloud vertical extent and the strength of the Na absorption feature centered around 590 nm. In all four cases, the differences between the laboratory and DDA phase curves compared to the TTHG phase curves do not exceed ±3 ppm.

---

[7] https://github.com/mglodge/Irregulator
[8] https://doi.org/10.5281/zenodo.14282625

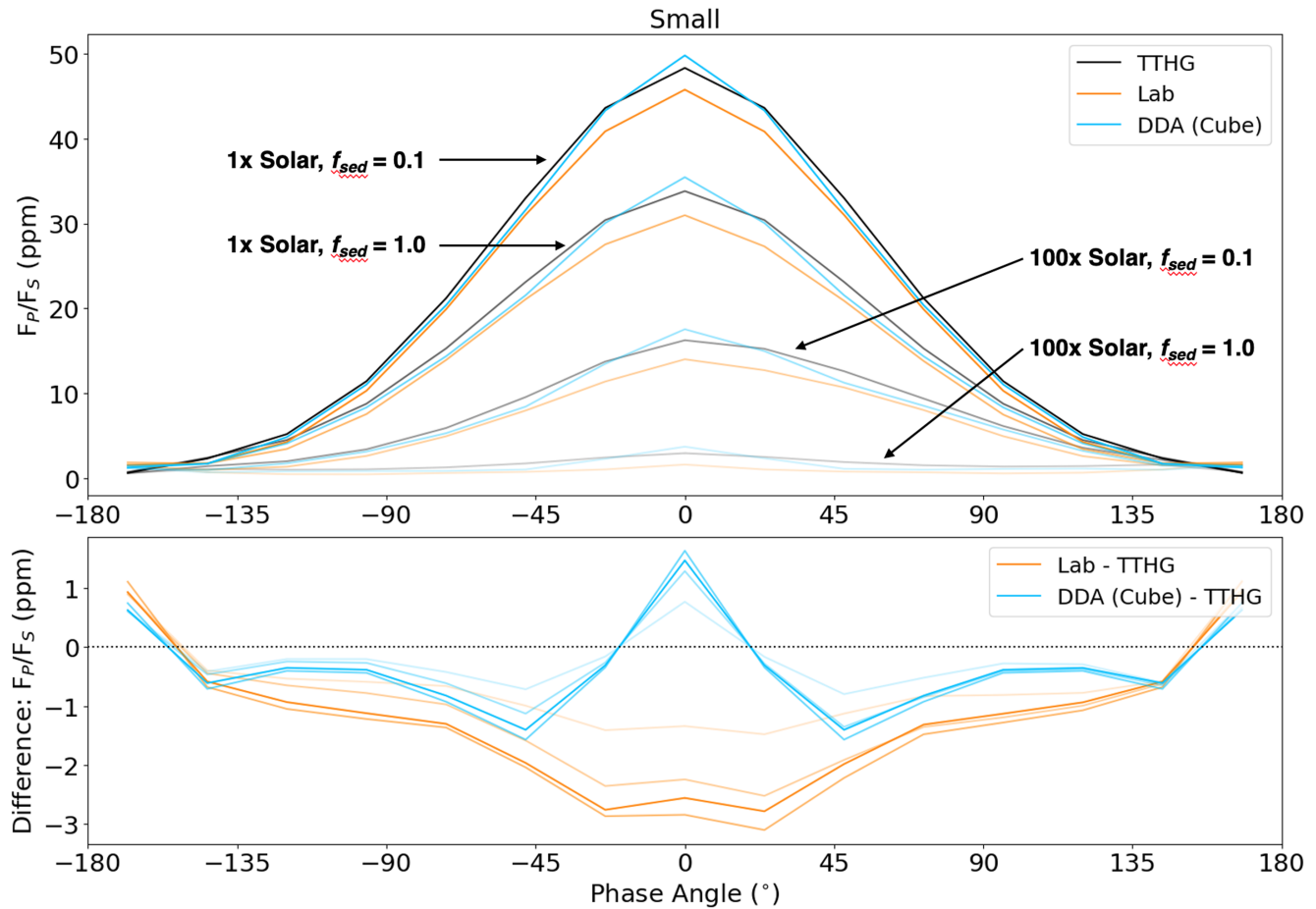

**Figure 9.** (Top) Reflected light phase curves computed from our GJ 1214b GCM at 532 nm for the small particle size distribution with the TTHG (black), laboratory (orange), or cubic DDA (sky blue) phase functions. Four different combinations of atmospheric metallicity and sedimentation efficiency are shown. Line opacities are used to differentiate the cases further, where high line opacities correspond to low atmospheric metallicity and low sedimentation efficiency, and low line opacities correspond to high metallicities and high sedimentation efficiencies. The differences in relative flux between the laboratory and TTHG phase functions (orange) and DDA and TTHG phase functions (sky blue) are shown in the bottom plot, with different line opacities indicating the different cases.

### ORCID iDs

Colin D. Hamill ⦿ https://orcid.org/0000-0002-9464-8494
Alexandria V. Johnson ⦿ https://orcid.org/0000-0002-6227-3835
Matt Lodge ⦿ https://orcid.org/0000-0002-9733-0617
Peter Gao ⦿ https://orcid.org/0000-0002-8518-9601
Rowan Nag ⦿ https://orcid.org/0009-0003-0371-296X
Natasha Batalha ⦿ https://orcid.org/0000-0003-1240-6844
Duncan A. Christie ⦿ https://orcid.org/0000-0002-4997-0847
Hannah R. Wakeford ⦿ https://orcid.org/0000-0003-4328-3867